\documentclass[conference,a4paper]{IEEEtran}
\IEEEoverridecommandlockouts

\usepackage{cite}
\usepackage{amsmath,amssymb,amsfonts}
\usepackage{algorithmic}
\usepackage{graphicx}
\usepackage{textcomp}
\usepackage{xcolor}
\usepackage{array}
\usepackage{multirow}
\usepackage{hyperref}

\newboolean{showcomments}
\setboolean{showcomments}{true}
\ifthenelse{\boolean{showcomments}}
{ \newcommand{\mynote}[3]{
        \fbox{\bfseries\sffamily\scriptsize#1}
        {\small$\blacktriangleright$\textsf{\emph{\color{#3}{#2}}}$\blacktriangleleft$}}}
{ \newcommand{\mynote}[3]{}}

\def\BibTeX{{\rm B\kern-.05em{\sc i\kern-.025em b}\kern-.08em
    T\kern-.1667em\lower.7ex\hbox{E}\kern-.125emX}}

\makeatletter
\newcommand{\linebreakand}{%
  \end{@IEEEauthorhalign}
  \hfill\mbox{}\par
  \mbox{}\hfill\begin{@IEEEauthorhalign}
}
\makeatother

\begin{document}

\title{Exploring the Power Consumption of ERC-4337 Bundling on Server Infrastructure\\
}

\let\oldIEEEauthorblockN\IEEEauthorblockN
\renewcommand{\IEEEauthorblockN}[1]{%
  \oldIEEEauthorblockN{\small #1}%
}

\let\oldIEEEauthorblockA\IEEEauthorblockA
\renewcommand{\IEEEauthorblockA}[1]{%
  \oldIEEEauthorblockA{\small #1}%
}

\author{
\IEEEauthorblockN{Claudiu-Nicu B\u{a}rbieru}
\IEEEauthorblockA{
\textit{Alexandru Ioan Cuza University}\\
Ia\c{s}i, Romania\\
claudiu.barbieru@gmail.com
}
\and
\IEEEauthorblockN{Oana-Otilia Captarencu}
\IEEEauthorblockA{
\textit{Alexandru Ioan Cuza University}\\
Ia\c{s}i, Romania\\
oana.captarencu@info.uaic.ro
}
\and
\IEEEauthorblockN{Pascal Felber}
\IEEEauthorblockA{
\textit{University of Neuch\^{a}tel}\\
Neuch\^{a}tel, Switzerland\\
pascal.felber@unine.ch
}
\and
\IEEEauthorblockN{Emanuel Onica}
\IEEEauthorblockA{
\textit{Alexandru Ioan Cuza University}\\
Ia\c{s}i, Romania\\
emanuel.onica@uaic.ro
}

\linebreakand

\IEEEauthorblockN{Valerio Schiavoni}
\IEEEauthorblockA{
\textit{University of Neuch\^{a}tel}\\
Neuch\^{a}tel, Switzerland\\
valerio.schiavoni@unine.ch
}
\and
\IEEEauthorblockN{Cosmin-Nicolae V\^{a}rlan}
\IEEEauthorblockA{
\textit{Alexandru Ioan Cuza University}\\
Ia\c{s}i, Romania\\
cosmin.varlan@info.uaic.ro
}
\and
\IEEEauthorblockN{Boris Verdecia Echarte}
\IEEEauthorblockA{
\textit{University of Neuch\^{a}tel}\\
Neuch\^{a}tel, Switzerland\\
boris.verdecia@unine.ch
}
\and

}

\maketitle
\begin{abstract}

The Ethereum ecosystem offers the most widespread support for Decentralised Applications (DApps), with over 120 networks currently documented as using variants of the Ethereum Virtual Machine (EVM) for core execution logic. 
The ERC-4337 standard provides convenient access to DApps 
for users who do not possess the required cryptocurrency or prefer flexible authorisation of DApp actions.
The infrastructure for ERC-4337 implementation can be deployed on any EVM network.
However, to aggregate user requests and submit them to the EVM network, ERC-4337 requires operating a bundler, an additional component.
Recently, the potential use of ERC-4337 was boosted by another standard, EIP-7702,
which can enable ERC-4337 utility for any existing user account. 
However, the use of the combined EIP-7702/ERC-4337 pattern remains considerably low at less than 6\% of EIP-7702 activity.  
We believe that this underutilisation is at least partly caused by the low number of existing bundling services and their popularity, currently fewer than 15 publicly documented.
Expanding the number of bundler nodes would benefit the ecosystem but would introduce operational costs, part of which is represented by the added nodes' power consumption.
In this paper, we provide a thorough view of the bundling power consumption aspect using server-grade infrastructure, evaluating both CPU and RAM consumption, as well as the overhead of container-based deployments specific to production environments.
\end{abstract}


\section{Introduction}
\label{sec:introduction}

The Ethereum~\cite{ethereum} ecosystem is currently the largest conglomerate of blockchain networks that provide support for decentralised applications (DApps).
At its core lies the main Ethereum network (L1), whose nodes rely on the Ethereum Virtual Machine (EVM) to execute smart contracts, small programs that implement the logic of functions changing the blockchain state. 
Users in the Ethereum ecosystem are normally identified by private-key controlled externally owned accounts (EOAs).
Transactions sent by EOAs can be either cryptocurrency transfers or, more generally, calls to smart contracts. 
The network nodes group these transactions into blocks, subject to consensus, and securely chain them together, providing decentralised, immutable storage guarantees.

Ethereum charges transaction fees to prevent network abuse and incentivise network participants.
Over time, the increase in the number of DApps backed by Ethereum smart contracts has led to network congestion and higher fees, a constant hurdle to onboarding new users. 
This resulted in a horizontal expansion via a multitude of secondary layer networks (L2).
These adapt or reuse the EVM, process user transactions similarly, and submit different kinds of commitments to the main network (transaction data batches, state summaries), thereby maintaining a trust anchor on Ethereum. 
This helped reduce congestion and fees, but new user onboarding remains a problem due to the complete lack of cryptocurrency for paying transaction fees or for interacting with Ethereum DApps~\cite{Chandra22}.

Account abstraction, standardised through ERC-4337~\cite{ERC4337}, introduces an independent accessibility layer that can be deployed on any EVM-based network.
The effective component of this layer that facilitates user access is a Smart Contract Account (SCA), which acts as a user account and can replace EOAs. 
Users can express interactions with target contracts through UserOperations (UserOps), which are subsequently executed by their SCA.
The SCA enables a variety of features that can enhance user experience, such as programmable transaction authorisation, fee sponsorship, transaction batching, and alternative authentication mechanisms, ultimately directing the UserOp to the target contract. 
To reach the SCA, UserOps are submitted to specialised nodes called \emph{bundlers}, which package multiple operations and forward them to the underlying EVM network. 
We describe the full ERC-4337 messaging and interaction path in Section~\ref{sec:background}.

Although convenient for users without an EOA, ERC-4337 smart contract accounts may add identity indirection and require existing EOA users to transfer assets to the SCA.
The introduction of EIP-7702~\cite{EIP7702} further expands the potential of ERC-4337 by allowing an EOA to have an execution logic linked to a deployed smart contract, which can be a SCA.
In this way, the delegated SCA logic executes in the context of the EOA, preserving its address and identity in calls to other contracts while having access to the assets held by the EOA.
Account abstraction functionality can therefore be offered to existing accounts without requiring users to migrate their funds to newly initialised SCAs. 
Nevertheless, the combined use of EIP-7702 and ERC-4337 remains limited, despite ERC-4337 being envisioned as the primary use case for EIP-7702 delegations~\cite{4337via7702} and recommended as such in official Ethereum best practices~\cite{Pectra7702}.
Statistics from December 2025 showed that seven months after the adoption of EIP-7702, less than 6\% of its monthly delegations used SCA facilities~\cite{EIP7702Review}.
Still, the end of 2025 marked the milestone of passing 1M weekly ERC-4337 UserOps in the context of EIP-7702 delegations, the figure rising to more than 2.5M in late June 2026~\cite{EIP7702Metrics}. 
This might seem like a solid increase, but the weekly transactions involving EIP-7702 delegated EOAs outside the ERC-4337 context exceeded 176M during the same period. 

We think that a contributing factor to the above situation is the limited number of publicly available bundling services that handle ERC-4337 UserOps.
Currently, there are no more than 15 officially documented bundlers on any of the major EVM networks~\cite{bundlersoverview}. 
Moreover, among these (excluding undocumented services), on each EVM network, only 3-4 bundlers account for more than 50\% of the bundling activity. 
Bundlers charge an additional fee for their service, which includes a validation of the UserOps, and grouping these in bundles.
Typically, this extra cost is computed as a percentage of the base transaction fee used on the EVM network.
Currently, popular bundler services add a 5\% overhead to this regular fee~\cite{Coombs2023,PimlicoFees} for UserOps.
This somehow undermines one of the main functions of ERC-4337: transaction sponsorship. 
Users who cannot cover regular transaction fees can wrap these as UserOps and rely on ERC-4337 mechanisms that allow the associated SCA’s fees to be sponsored by a third party. 
Potential sponsors, however, may incur higher costs than they would for regular transactions.
We believe that increasing the number of bundlers could foster wider adoption of ERC-4337, reduce costs through stronger market competition, and overall enhance the availability and decentralisation of ERC-4337 services.

An expansion of the number of bundlers must, however, account for complete operational costs, including energy consumption. 
Power usage has long been a sensitive concern for blockchain networks due to their computational demands. 
Ethereum's transition from Proof-of-Work to Proof-of-Stake consensus marked a significant reduction from an annual estimation of 21 TWh to 0.0026 TWh in its energy expenditure, an aspect that receives constant scrutiny~\cite{EthereumEnergy}.
Recent work~\cite{arusoaie2025} advocated co-hosting a bundler service alongside a regular Ethereum client implementation on the same node, which would naturally increase the number of bundlers. 
A bundler performs repeated simulations for validation purposes, transaction cost estimation requests, and even debug traces of UserOps, all of which are executed via RPC requests to an EVM client on the network.
Therefore, co-hosting bundlers with EVM implementations would reduce communication overhead with the EVM network and also cut the costs of running separate bundler nodes.
While energy costs may be of low importance to an independent operator running a limited number of bundler nodes, this cost significance is worth considering at the broader level of the ecosystem.
There is no exact number of Ethereum nodes due to variable network churn, but estimates currently put the number at around 11K~\cite{EthereumNodeTracker}. 
The majority of L2 EVM-based networks do not use a decentralised consensus, which presumably makes their node counts much lower, but, altogether, even harder to estimate in the absence of official records. 
Furthermore, the L2 number is constantly expanding, recent studies~\cite{ishmaev2025} counting 129 networks. 
Running a bundler instance on each EVM node would likely be unnecessary, but determining the optimal bundler-to-network ratio is a complex problem that requires a more advanced scalability study that accounts for real-time variations such as network churn and load. 
Intrinsically, this makes determining a bundler's power consumption a potentially important factor in such a volatile context. 
Evaluating the energy implications of emerging blockchain infrastructure is necessary before its deployment reaches ecosystem scale.

Previous early results~\cite{arusoaie2025} indicate that operating an ERC-4337 bundler does not inflict a substantial average power consumption overhead on an EVM client, and should not represent a concern at the level of an individual host. 
These estimates, however, were limited to empirical evaluations of CPU power consumption on mobile hardware and cannot precisely characterise the impact of deploying a bundler within the Ethereum ecosystem. 
They do not include measurements on production environments, which commonly rely on containerisation and consider server-grade infrastructure.
In this paper, we investigate the power-consumption implications in a realistic setting that would fit a broader deployment of ERC-4337 bundlers. 
We evaluate the average power consumption of both the CPU and memory across several server-grade configurations and quantify the power use of container-based deployments that are representative of production environments. 
We use SmartWatts~\cite{Fieni2020SmartWatts} for our measurements, a well-established power-consumption estimation software whose accuracy has been confirmed in scientific evaluations~\cite{Khan2016}. 

\subsubsection*{Paper Organisation}
In Section~\ref{sec:background}, we provide additional background on the ERC-4337 communication flow and details on using SmartWatts for measuring the average power consumption.
In Section~\ref{sec:setup}, we elaborate on a series of details that provide specificity to the evaluation of the bundler component in a production environment, and we also lay the groundwork for future, more in-depth investigation. 
We follow in section~\ref{sec:evaluation} with the description of the experimental setup, the conducted evaluation and the results we obtained. 
We conclude in Section~\ref{sec:conclusion}.

\section{Background}
\label{sec:background}

\begin{figure*}[ht]
  \centering
  \includegraphics[width=\textwidth]{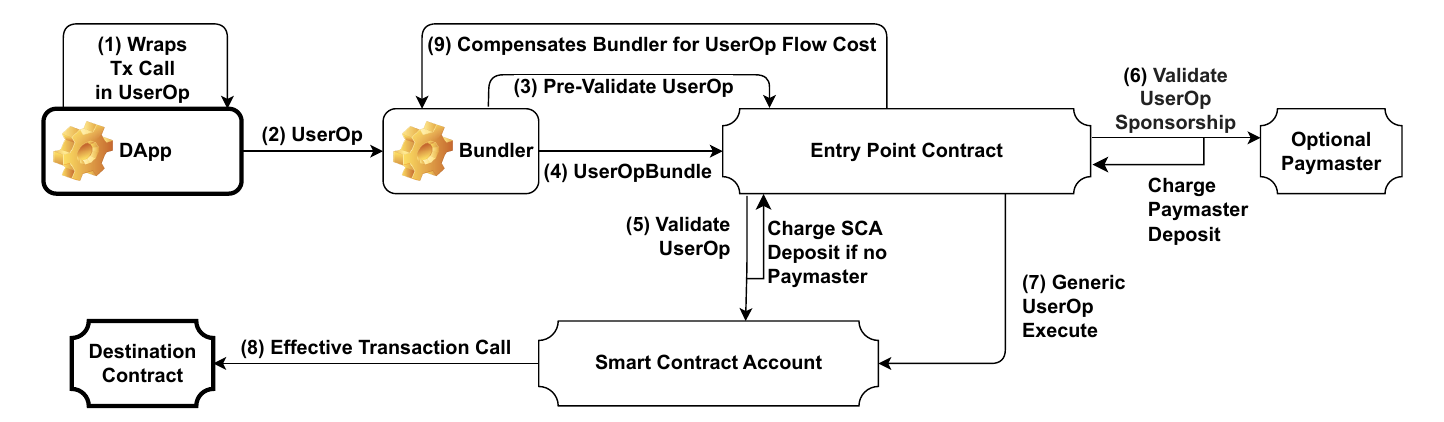}
  \caption{\label{figuserop}Transaction communication path in ERC-4337}
\end{figure*}

In what follows, we start by discussing the ERC-4337 UserOp path and the involvement of a bundler in its processing. 
Then we continue with background on conducting power consumption measurements.

\subsection{The ERC-4337 UserOps Workflow}

Figure~\ref{figuserop} illustrates the communication flow involving the bundler in ERC-4337. The other participants are the user’s smart contract account (SCA), the destination contract with which the user intends to interact, an EntryPoint contract on the Ethereum network, and an optional Paymaster - a contract willing to sponsor the user’s transaction.
First, a DApp constructs a UserOperation (UserOp) message. 
The UserOp identifies the SCA as the sender, encodes the call to the destination contract into the \emph{callData} field, and includes the custom signature required by the SCA’s validation logic (1).
The UserOp is then submitted to the bundler (2). 
The bundler performs an initial validation of the UserOps received (3) and aggregates multiple valid operations into a bundle. 
Subsequently, it submits the bundle to the EntryPoint contract through a transaction (4). 
The bundler must initially pay the gas cost of this transaction.

The EntryPoint then validates each UserOp, interacting with the SCA associated with the UserOp initiator (5). 
The UserOp can designate a third-party paymaster to sponsor the transaction fee.
In this case, the EntryPoint also validates the UserOp through the Paymaster (6) and charges the execution costs to either the SCA or the Paymaster, depending on whether sponsorship is used.

Then, the EntryPoint invokes a generic execution function in the SCA (7), forwarding the payload specified in the UserOp’s callData. 
This causes the SCA to perform a regular call to the destination contract (8).
Finally, the EntryPoint compensates the bundler for the gas spent on the bundle transaction (9).

We note that step (2) in Figure \ref{figuserop} can be more complex in practice, as some bundlers may share a common UserOp mempool. 
In this case, the UserOp may be further propagated among bundlers, potentially increasing the overhead of the bundler process.
However, not all bundlers support this functionality. 
Also, this UserOp distribution relies on gossip-based propagation rather than broadcast, meaning that each bundler forwards the UserOp only to a limited set of peer nodes, with less impact on power consumption.

Intuitively, the computational overhead introduced by the bundler process is expected to result primarily from the processing of UserOps. 
Steps (3) and (4) are performed via remote procedure calls (RPC).
These use the JSON-RPC protocol supported by the EVM clients on the blockchain network. 
Hence, the bundler-side overhead associated with these steps is largely limited to preparing, issuing, and processing the corresponding RPC requests and responses.
A substantial part of the workflow (steps (5)–(8)) involves only the blockchain nodes that handle the execution of the relevant smart contracts. 
Therefore, these operations do not contribute directly to the computational workload of the bundler. 
Similarly, step (9) is a payment transaction, also executed on the blockchain nodes.

We notice that the ERC-4337 flow introduces a more substantial overhead on the blockchain node running the EVM that executes the EntryPoint transaction. 
However, this overhead is inherent to the on-chain execution of the ERC-4337 workflow and is therefore independent of where the bundler node is deployed.
 
\subsection{Power consumption estimation}
A computer’s power consumption is strongly affected by the software executed on the system~\cite{Colmant2018,Acar2016}. 
In general-purpose computing, the CPU is a major source of variation in power drawn, which correlates with the consumption of the entire system ~\cite{Colmant2018,Rivoire2008HighLevel}.

Although various physical power meters can be used to measure the total power drawn by a node, isolation the power consumption of a specific software process is more challenging. 
A coarse estimate may be obtained by averaging the difference between multiple executions performed with and without the target software. 
However, this approach cannot isolate several factors that influence active power consumption: background activity from software dependencies and other processes, CPU voltage and frequency changes introduced by Dynamic Voltage and Frequency Scaling (DVFS), cache misses associated with the execution context, and transitions into low-power C-states during idle periods.
These limitations have motivated the development of power models~\cite{Colmant2018,Do2009pTop}. Many combine system-level measurements with machine-learning techniques to provide more accurate estimates of CPU power consumption.

Our study relies on SmartWatts~\cite{Fieni2020SmartWatts}, an open-source software that estimates the power consumption of software processes, containers, virtual machines, or applications running on a physical server. 
This is implemented as part of the PowerAPI framework~\cite{powerapiorg,Colmant2018}, which leverages Running Average Power Limit (RAPL)~\cite{IntelSDM}.
RAPL is an interface developed by Intel that permits capturing the active power consumption of several hardware domains. 
This interface was introduced in Intel processors with the Sandy Bridge architecture and later adopted by AMD in the Zen architecture. 
RAPL provides energy measurements for the CPU package, separate processor cores, shared caches, DRAM and memory controllers, integrated graphics, depending on the processor architecture and the domains supported by the platform.
Although RAPL enables software-based measurements of CPU power consumption, it provides only a global estimate whose accuracy may be limited~\cite{Colmant2015}.

To improve measurement precision, SmartWatts complements RAPL data with hardware performance counters (HwPCs) collected through the CPU’s performance monitoring units (PMUs). 
The monitored events include the APERF/MPERF ratio, which reflects the relationship between the effective and nominal core frequencies and captures turbo-boost and power-saving behavior, as well as unhalted core cycles, last-level cache misses that may trigger main-memory accesses, and completed instructions. 

\section{Bundler Power Consumption:\\Production Environment Considerations}
\label{sec:setup}

In this section, we outline specific details about running a bundler in a production environment and its implications for power consumption.  
We begin with a concise set of hardware considerations, followed by the essential software necessities and their potential relation to power consumption. 
We continue with brief considerations on power consumption measurement and conclude the section by pointing to advanced details of the bundling process that may be explored in future work. 

\subsection{Hardware Setup Considerations}

To our knowledge, the only prior work that considered bundler power-consumption  is~\cite{arusoaie2025}. 
This empirically estimated the overhead of the bundler running on an EVM client node, using a mobile processor that exposes the RAPL interface, an Intel i7-10870H. 
From a production environment perspective, a more accurate evaluation should consider server-grade CPUs, e.g., the Intel Xeon family or AMD EPYC. 

Server-grade processors typically have a higher baseline idle power consumption due to larger caches and, overall, more complex I/O and uncore subsystems.
Besides that, larger caches may reduce DRAM accesses for cache-resident workloads, shifting consumption from DRAM toward CPU activity.
On the other hand, server-grade setups typically come with much greater memory capacity and, consequently, higher idle power consumption for DRAM as well.
This also prompts consideration of the memory power consumption measurement, which, to our knowledge, has not been performed before in bundler contexts. 
Finally, although the Thermal Design Power (TDP) is related to dissipated heat rather than power consumption, it still indicates a possible power envelope that is considerably higher for server-grade CPUs. 

Although maintaining an empirical level for our observations, in our current work, we focus exclusively on server-grade hardware, which we consider considerably more relevant for the overall context of bundler node expansion.

\subsection{Baseline Software Considerations}

As discussed in the previous section, all ERC-4337 bundlers follow a common workflow. 
However, implementations may differ in specific internal stages, such as the ordering and grouping of UserOps when constructing bundle transactions, as well as validation practices, including simulation methods and the frequency of checks.
Therefore, different bundler software can have varying impacts on power consumption. 

Currently, aside from the reference implementation~\cite{infinitismbundler}, the diversity of bundler clients is relatively limited.
Pimlico (Alto)~\cite{altobundler}, the bundler under scrutiny in the previous work~\cite{arusoaie2025}, remains in the top three by total UserOps load handled in the EVM ecosystem. 
This bundler holds the first position on the main Ethereum network~\cite{EthereumBundlers} and a couple of other EVM networks. 
Other top bundler services include Alchemy~\cite{alchemybundler}, Thirdweb~\cite{thirdweb} and 
Coinbase~\cite{coinbase}.
These four bundler providers concentrate more than 50\% of the bundling activity~\cite{bundlersoverview}.
Notably, Pimlico and Alchemy open-source their respective bundler implementations: Alto~\cite{altobundler} and Rundler~\cite{alchemybundler}, respectively. 
By contrast, Thirdweb and Coinbase publish integration tooling, client libraries, and orchestration examples but do not currently disclose the server-side bundler implementations used by their production services.

Besides differences in bundler software, power consumption variations are also amplified by the programming language used for bundlers' implementations.
Notable overheads that pass 20x have been observed between different languages, such as TypeScript, used by Alto, over Rust, used by Rundler~\cite{Pereira2017}. 
If we consider running the bundler software on the same node as the EVM, the overall power consumption also depends on the EVM client node implementation due to frequent interactions with the bundler. 
The EVM node implementations also vary.
Therefore, to isolate the bundler's effective overhead, previous work~\cite{arusoaie2025} uses Anvil~\cite{anvil}, a development node simulator.
This provides a reference EVM implementation focused on the baseline operations of executing contracts and forming the chained blocks, rather than on ensuring consensus or other complementary operations. 

Finally, power consumption is affected by the logic of the executed transactions.
While this logic is independent of the bundler implementation, specific workloads can indicate how significant power consumption may be.
Previous work~\cite{arusoaie2025} has considered token transfers in particular.
These are among the most popular transactions in the EVM ecosystem, with ERC-20~\cite{ERC20} reaching almost 2 billion trades by 2025~\cite{somin2025cryptoasset}.
Moreover, Uniswap, which heavily uses ERC-20 tokens, is currently the decentralised exchange protocol with the widest spread, with more than 740 forks~\cite{defillama-uniswap}.
Therefore, we consider ERC-20 a proper use-case indicator. 

\subsection{Power Measurement Considerations}

The aforementioned prior work~\cite{arusoaie2025} used the SmartWatts~\cite{Fieni2020SmartWatts} component of the PowerAPI toolkit~\cite{powerapiorg} to estimate power consumption, taking as input the CPU exported RAPL data, complemented by HwPC. 
This data is mapped to processes using the separation provided by Linux control groups (cgroups). 
First, SmartWatts estimates the system’s idle power baseline and then applies a regression algorithm to obtain the active power consumption split per process group. 
The power consumption estimation model selects HwPC events based on their correlation with RAPL measurements and periodically recalibrates to maintain an error of less than 3.5\%.~\cite{Fieni2020SmartWatts}

We consider a similar approach using SmartWatts, which allows us to maintain consistency with previous measurements, even though those were obtained on a completely different hardware setup. 
Still, since we use similar, albeit updated, software choices as discussed above, we can draw some baseline comparison with former results. 

This former work isolated the EVM client (Anvil) and bundler (Alto) processes by running them in transient scope units, which function as temporary cgroups, enabling SmartWatts to capture their distinct average power consumption. 
This can provide a perspective on native runs on bare-metal, server-grade infrastructure.
However, in production environments, it is often preferred to run services in containers for isolation, portability, simplified dependency management, and easier scaling and rollback. 
In the current work, we focus on such a setting, using Docker containers.
However, Docker containers could trigger potential power consumption overhead, which we investigate in our evaluation.  

\subsection{Advanced Bundling Considerations}

Our current study evaluates a bundler's power consumption overhead in the baseline environment described above, centred on token use, which is representative of typical blockchain workloads. 
However, ERC-4337 provides context for more complex use cases, such as sponsoring transactions through a paymaster contract and enforcing custom authentication. 
The impact of power consumption in such cases is mainly on the blockchain's EVM node, which executes the transactions.
Still, several aspects may be important to discuss in the context of bundling. 
While outside the scope of the current work, we consider the following for potential future exploration.

\subsubsection*{Transaction Costs Estimation Traffic}
A bundler may repeatedly execute simulation RPC calls to estimate UserOp costs (e.g., when servicing \textit{eth\_estimateUserOperationGas} requests). 
This may involve simulating the transaction's inner execution, incurring additional computational costs. 
Moreover, traffic also adds communication costs, which might be reflected in the bundler's power consumption. 

\subsubsection*{Smart Contract Account Types}
For the baseline in our study, we consider the reference SCA implementation (SimpleAccount)~\cite{infinitism}, with the default typical Ethereum ECDSA signing for submitted UserOps. 
However, SCA implementations can opt for more complex UserOp signature validation, including multisigs, session keys, passkeys and other custom mechanisms. 
The signature verification can therefore produce a computational overhead.
This is a generic factor, independent of the transaction logic, that can increase the EVM node's power consumption during transaction validation.

\subsubsection*{Validation Complexity}
In addition to signature verification, the pre-validation of UserOps performed by the bundler may affect power consumption, depending on the complexity of the transaction validation logic and the contract call depth. 
A proposed standard, ERC-7562~\cite{ERC7562}, introduces validation-scope rules intended to protect bundlers against DoS attacks.
To enforce these rules, bundlers trace the validation phase and detect prohibited opcodes, restricted storage or code accesses, and other disallowed dependencies.
Regardless of the ERC-7562 tracing, a bundler will still trigger a simulation of the validation logic, which may increase the connected EVM node's computations and power consumption.

\section{Evaluation}
\label{sec:evaluation}

We considered three production server configurations for our evaluation, as follows:

\subsubsection*{Intel Xeon Gold (IXG)} The first server is equipped with a 16-core (32 threads) 3rd Gen Intel Xeon Gold 6326 CPU, 185W TDP, running from a baseline frequency of 2.90 GHz up to a maximum turbo of 3.50 GHz. 
The server has 64 GB DDR4 memory installed, running at 3200 MT/s and uses a 512 GB NVMe SSD.
The operating system on this configuration is Ubuntu 22.04.5, using the Linux 5.15.0-186-generic kernel.

\subsubsection*{AMD EPYC (AMD)} The second server is equipped with a 16-core (32 threads) AMD EPYC 7302P CPU, 155W TDP, with a baseline frequency of 3.00 GHz and a maximum boost of 3.50 GHz. 
This server has 32 GB DDR4 running at 3200 MT/s and uses a 480 GB SATA SSD.
The operating system on this configuration is similar to the IXG.

\begin{table*}[ht]
  \caption{Weighted Average of Watts Power Consumption on Intel Xeon Gold}
  \label{tab:ixg}
\renewcommand{\arraystretch}{1.3}
\centering
\begin{tabular}{| m{0.11\linewidth} || m{0.10\linewidth} | m{0.10\linewidth} | m{0.10\linewidth} || m{0.10\linewidth} | m{0.10\linewidth} | m{0.10\linewidth} |} 
 \hline
 Test Config & Alto CPU (Docker/Native) & Anvil CPU (Docker/Native) & Global CPU (Docker/Native) & Alto DRAM (Docker/Native) & Anvil DRAM (Docker/Native) & Global DRAM (Docker/Native) \\ 
 \hline\hline
System at Rest & 1.23 / 0.91 & 1.67 / 0.94	& 36.42 / 34.08 & 0.38 /	0.18 & 0.32 / 0.13	& 8.49 / 7.46 \\\hline\hline
\multicolumn{7}{|c|}{Variation of Throttling per UserOp with 100 UserOps Burst Load per Round and 12s Blocks Frequency} \\\hline
25ms Throttling & 14.92 / 15.15 & 3.30 / 3.58 & 37.09 / 36.94 &	1.77 /	2.74	& 0.27	/ 0.68 & 7.38 / 7.27 \\\hline
50ms Throttling & 11.13	/ 10.66 & 4.61 / 3.41 &	36.38 /	37.27 &	2.24 / 2.04 &	0.91 /	0.67 &	7.61 /	7.59 \\\hline
100ms Throttling & 5.83	/ 6.64 & 7.05 /	5.14 &	37.60	/ 37.65 & 1.51 / 1.71 &	1.26 / 0.92 & 7.82 /	8.01 \\\hline\hline
\multicolumn{7}{|c|}{Variation of UserOps Burst Load per Round with 25 ms Throttling per UserOp and 12s Blocks Frequency} \\\hline
75 UserOp Burst & 13.08	/ 12.56	 & 4.37 /	2.89 &	36.21 /	36.06 & 2.55 /	2.47 &	0.89 / 0.61	 & 7.31 /	7.26 \\\hline
50 UserOp Burst & 6.85 / 8.99 &	5.31 /	3.14 &	35.19 /	34.20 & 1.12 /	0.86 &	1.08	/ 0.25 &	7.37 /	7.18 \\\hline
25 UserOp Burst & 4.60 / 6.56 &	5.48 /	3.57 &	35.47 /	35.84 & 1.10 /	1.78 &	1.06 / 0.64	& 7.44 /	7.67  \\\hline\hline
\multicolumn{7}{|c|}{Variation of Block Frequency with 100 UserOps Burst per Round and 25 ms Throttling per UserOp} \\\hline
2s Block & 25.22 /	27.91	& 12.01 /	7.31 &	44.96 /	44.27 & 4.26 /	4.90 &	2.06 /1.19 &	7.96 /	7.99 \\\hline
6s Block & 19.37 /19.66 &	6.37 /	4.28& 39.27 /	39.02 & 3.44 /	3.46 &	1.20 /	0.78 &	7.47 / 7.35  \\\hline
\end{tabular}
\vspace{-10pt}
\end{table*}

\begin{table*}[ht]
  \caption{Weighted Average of Watts Power Consumption on Intel Xeon Silver}
  \label{tab:ixs}
\renewcommand{\arraystretch}{1.3}
\centering
\begin{tabular}{| m{0.11\linewidth} || m{0.10\linewidth} | m{0.10\linewidth} | m{0.10\linewidth} || m{0.10\linewidth} | m{0.10\linewidth} | m{0.10\linewidth} |} 
 \hline
 Test Config & Alto CPU (Docker/Native) & Anvil CPU (Docker/Native) & Global CPU (Docker/Native) & Alto DRAM (Docker/Native) & Anvil DRAM (Docker/Native) & Global DRAM (Docker/Native) \\ 
 \hline\hline
System at Rest & 2.3 / 2.01 &	8.98 /	9.9 &	45.5 /	45.52 & 0.28 /	0.27 &	1.08 /	1.15 &	5.48 /	5.42  \\\hline\hline
\multicolumn{7}{|c|}{Variation of Throttling per UserOp with 100 UserOps Burst Load per Round and 12s Blocks Frequency} \\\hline
25ms Throttling & 17.36 / 16.16 &	9.07 /	12.78 &	55.85 /	59.78 & 2.03 /	2.40 &	1.04 /	1.80 &	6.52 /	7.94 \\\hline
50ms Throttling & 12.57 / 12.79 &	11.05 /	13.13 &	56.81 /	59.77 & 3.27 /	1.64 &	2.60 /	1.49 &	9.68 /	7.03 \\\hline
100ms Throttling & 7.97	/ 5.86 & 11.85 / 13.46 & 60.17 / 58.30 & 0.88 /	0.69 &	3.41  /	1.73 &	8.91 /	7.34 \\\hline\hline
\multicolumn{7}{|c|}{Variation of UserOps Burst Load per Round with 25 ms Throttling per UserOp and 12s Blocks Frequency} \\\hline
75 UserOp Burst & 14.86	/ 14.89 &	10.61 /	12.10 &	57.57 /	56.04 & 3.39 /	1.80 &	2.96 /	1.44 &	10.32 /	6.71 \\\hline
50 UserOp Burst & 11.28	/ 10.78 &	12.77 /	12.51 &	59.35 /	56.14 & 1.40 /	1.44 &	1.51 /	1.41 &	7.22 /	6.48 \\\hline
25 UserOp Burst & 7.96 / 7.39 &	12.37 /	13.43 &	59.35 /	52.79 & 0.83 /	0.91 &	1.47 /	1.59 &	7.16 /	6.44 \\\hline\hline
\multicolumn{7}{|c|}{Variation of Block Frequency with 100 UserOps Burst per Round and 25 ms Throttling per UserOp} \\\hline
2s Block & 23.49 /	23.29 &	20.31 /	19.90 &	62.26 /	62.15 & 2.86 /	2.87 &	2.21 /	2.12 &	7.26 /	7.22  \\\hline
6s Block & 22.01 /	20.74 &	12.84 /	12.22 &	60.25 /	56.79 & 3.70 /	2.50 &	2.37 /	1.33 &	9.29 /	6.52 \\\hline
\end{tabular}
\vspace{-10pt}
\end{table*}

\begin{table}[t]
 \caption{Weighted Average of Watts Consumption on AMD EPYC}
  \label{tab:amd}
\renewcommand{\arraystretch}{1.3}
\centering
\begin{tabular}{| m{0.23\linewidth} || m{0.19\linewidth} | m{0.19\linewidth} | m{0.19\linewidth} |} 
 \hline
 Test Config & Alto CPU (Docker/Native) & Anvil CPU (Docker/Native) & Global CPU (Docker/Native) \\ 
 \hline\hline
System at Rest & 0.84 /	0.88 &	6.52 /	1.64 &	41.50 /	41.18   \\\hline\hline
\multicolumn{4}{|c|}{Variation of Throttling with 100 UserOps Bursts and Blocks each 12s} \\\hline
25ms Throttling & 16.44 / 16.78 &	6.83 /	6.60 &	43.96 /	43.70\\\hline
50ms Throttling & 13.09 /	13.39 &	7.31 /	6.28 &	44.50 /	47.46 \\\hline
100ms Throttling & 6.01	/ 7.65 &	8.28 /	5.88 &	46.32 /	46.95\\\hline\hline
\multicolumn{4}{|c|}{Variation of UserOps Bursts with 25 ms Throttling and Blocks each 12s} \\\hline
75 UserOp Burst & 15.69 / 14.85 &	9.92 /	6.32 &	43.59 /	42.74 \\\hline
50 UserOp Burst &  12.44 /	13.13 &	9.76 /	6.09 &	42.43 /	41.41\\\hline
25 UserOp Burst & 7.58 / 9.12 &	9.05 /	4.65 &	41.17 /	40.43 \\\hline\hline
\multicolumn{4}{|c|}{Variation of Block Freq. with 100 UserOps Bursts and 25 ms Throttling} \\\hline
2s Block & 27.33 /	28.49 &	13.25 /	9.09 &	49.25 /	49.05 \\\hline
6s Block & 21.50 /	23.75 &	9.91 /	7.07 &	45.99 /	46.23 \\\hline
\end{tabular}
\end{table}

\subsubsection*{Intel Xeon Silver (IXS)} The third server is equipped with a 12-core (24 threads) 5th Gen Intel Xeon Silver 4510 CPU, 150W TDP, with a baseline frequency of 2.40 GHz that goes up to a max turbo of 4.10 GHz.
This server has 128 GB DDR5 running at 5600 MT/s and uses a 512 GB SATA SSD set in RAID 1 mode. 
The operating system on this server is Debian 13 as the base for the Proxmox 9.1 virtualisation environment, using the specific Linux 6.7.13-4-pve kernel build.

For our software setup, we opted for the configuration described in the previous section.
We used Anvil v1.7.1 as an EVM simulator and Alto as a bundler.
To enable the ERC-4337 flow, we deployed EntryPoint v0.7, which is currently the most used in the Ethereum ecosystem~\cite{entrypoint7}. 
For the workload contents, we took a similar approach as previous work~\cite{arusoaie2025}.
We considered bursts of ERC-20 transfer transactions, processed as UserOps via a set of distinct deployed SCAs that use the reference SimpleAccount implementation~\cite{infinitism}.
Unlike the aforementioned previous work, which performs an empirical evaluation locally confined to an end-user machine, we considered a production-oriented simulation in which we fed the UserOps from an external client machine to each server-hosted bundler. 
We prepared our Alto and Anvil services setup in two variants: one that runs the processes in Docker containers, which is more appropriate for a production scenario where isolation and portability are preferred, and another where the processes run natively on the server machines.

\begin{figure*}[ht]
  \centering
  \includegraphics[width=\textwidth]{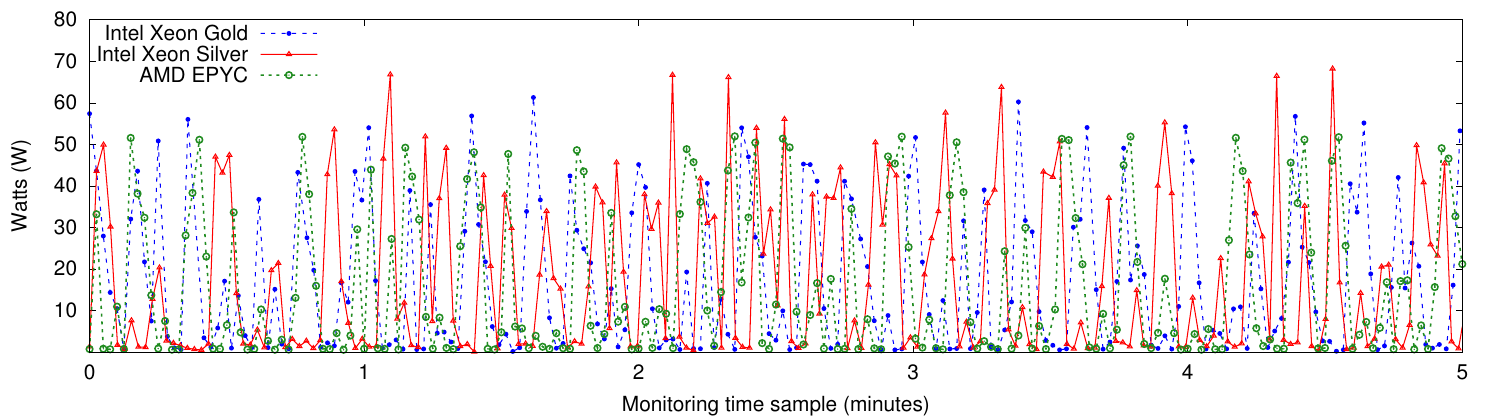}
  \caption{\label{fig:alto}Power Consumption Sampling of Alto (100 UserOps bursts, 25ms throttling per UserOp, 12s block frequency)}
\vspace{-5pt}
\end{figure*}

Our evaluation focus was threefold. 
First, as in the previous study~\cite{arusoaie2025}, we were interested in the overhead that the bundler brings to power consumption when running on an effective production node along with an EVM client.
Second, we were interested in observing if production-oriented Docker setup brings a significant overhead compared to native runs.
Finally, we wanted to observe how the different hardware architectures compare in the power consumption for Alto and Anvil. 

We leveraged SmartWatts to estimate the active power consumption scoped to the Alto and Anvil runs, supplemented by the global system load.
The Intel server configurations expose both CPU and DRAM RAPL counters, allowing us to measure both values.
The AMD configuration is limited to CPU data, similar to the evaluation context of the previous work~\cite{arusoaie2025}.
The energy readings were captured at a 2Hz sampling interval, the default for SmartWatts, which can slightly vary depending on the hardware response. 
These sampling time variations were captured in the final results weighted average, to which SmartWatts estimates an error in the range of 2 Watts.

For most of our tests, we followed a methodology set in previous work~\cite{arusoaie2025}. 
We used UserOps burst loads of sizes from 25 to 100, addressed to different SCAs.
We introduced various throttling between each UserOp sending.
This simulates the variability of a real scenario, when the bundler could receive loads of requests triggered by certain events, but not all simultaneously. 
Finally, we varied the produced block frequency from 12 seconds, which is the Ethereum L1 standard, to lower values: 6 and 2 seconds, which are more specific for L2 EVM networks. 
The duration of the tests was set in the number of burst rounds, ranging from 50 to 300. 
One UserOp burst is typically processed in one or two blocks.
Therefore, given the block intervals, the tests ran for at least 10 minutes. 

We summarise the weighted average power consumption across the various tests for the three configurations in Table~\ref{tab:ixg} (IXG), Table~\ref{tab:ixs} (IXS), and Table~\ref{tab:amd} (AMD).
Overall, we observe the expected results: power increases with lower throttling, higher UserOp loads, and shorter block intervals. 

A first notable difference compared to previous empirical evaluations in~\cite{arusoaie2025} is that on all configurations Alto typically has a higher power consumption profile than the Anvil EVM client. 
The exceptions when Anvil tends to bypass Alto are the idle measurements and the tests with high throttling (100ms) or low UserOp burst load (25) at longer block intervals.
This is explained by Anvil still running block production even with empty blocks. 
We believe that Alto dominates Anvil consumption in the rest of the tests because of the complexity of the ERC-4337 workflow compared to Anvil, which is a simulation client rather than a full EVM node. 
Further, Anvil was optimised starting from version 1.7~\cite{anvil17}.
The important take-away remains the effective Alto overhead. 
While the average CPU power observed in previous mobile tests was below 4W, in production environments it can exceed 25W in the most demanding tests, with occasional spikes above 70W.

For Docker-based and native runs, aside from a few outliers we attribute to concurrent operating system activity, their power consumption is rather similar. 
For tests using a 12-second block frequency, the average differences between Docker and native Alto are 0.69W for both Intel configurations and 0.61W for AMD. 

Concerning the different hardware architectures, we observe that Intel Xeon Silver is the most power-demanding. 
Figure~\ref{fig:alto} displays a sample of the Alto power consumption captured on the test configurations.
The difference between IXS and IXG can be explained by the fact that Xeon Silver uses a newer-generation processor, with a higher maximum turbo frequency. 
Given the burst characteristics of the workload, we noticed that the IXS CPU often reaches frequencies of \~4GHz, which is higher than the maximum of the other processors. 
Also, the AMD CPU has a slightly higher base frequency than the IXG, resulting in a similar effect. 
Moreover, the Xeon Silver features 80 newer DDR5 PCIe 5.0 lanes, whereas the 6326 uses 64 DDR4 PCIe 4.0 lanes, which also contributes to the increase in power consumption.
Regarding DRAM, we remark that this is considerably less significant than the CPU in the overall power consumption. 
For the tests that use a 12-second block frequency, Alto DRAM consumption does not exceed 20\% of the CPU consumption, generally remaining below 3W.

\section{Conclusion}
\label{sec:conclusion}

In this paper, we measured the power costs of ERC-4337 bundling on production server infrastructure. 
To our knowledge, this is the first evaluation of this kind, with previous work limited to estimations on mobile processors. 
We tested multiple configurations of UserOp workloads on three different representative server configurations: Intel Xeon Gold, Intel Xeon Silver and AMD EPYC.   
The highest weighted averages of the bundler were in the range of 25W, compared to 4W on the mobile platforms. 
This is expected for a production platform and can still be regarded as a low value in terms of costs, given that server-grade CPUs have considerably higher TDP (e.g., 185W for the Xeon Gold compared to \~50W or lower on mobile processors).
Our findings confirm the feasibility of co-hosting the bundler with an EVM node from an energy-efficiency perspective.
The correlated results of our study that show non-significant differences in running bundler services in containers also strengthen the idea of a colocated service.  
Future research can investigate the advanced bundling considerations that we identified, as well as the power consumption of other bundlers.

\section*{Acknowledgment}
This work was supported by CHIST-ERA-23-SMARTC-01-SCEAL project, through a grant from the Romanian Ministry of Research, Innovation and Digitization, CNCS/CCCDI - UEFISCDI, project number 86/2025 ERANET-CHISTERA-IV-SCEAL, within PNCDI IV and from the Swiss National Science Foundation under Project 20CH-1\_229451.

{
\bibliographystyle{IEEEtran}
\bibliography{biblio}
}

\end{document}